\documentclass[conference]{IEEEtran}
 
\usepackage{cite}

\usepackage{gensymb}
\usepackage{soul}
\usepackage{lipsum}
\usepackage{multicol}

\usepackage{booktabs}
\usepackage{array}

\ifCLASSINFOpdf
\usepackage[pdftex]{graphicx}

\else
   
\fi

\usepackage[cmex10]{amsmath}
 
\usepackage[font=small,labelfont=bf,
   justification=justified,
   format=plain]{caption}

\usepackage{algorithmic}
\usepackage{subfig}

\usepackage{tikzpagenodes}
\usepackage{lipsum}

\usepackage{mdwmath}
\usepackage{mdwtab}
 
\usepackage[justification=centering]{caption}

\usepackage{placeins}

\begin{document}
%
\title{An Accurate Smartphone Battery Parameter Calibration Using Unscented Kalman Filter} 
\author{\IEEEauthorblockN{Chalukya Bhat$^{1}$, \textit{Student Member, IEEE}, Aniruddh Herle$^{2}$, \textit{Student Member, IEEE},   Janamejaya Channegowda$^3$, \\ \textit{Member, IEEE}, Kali Naraharisetti$^4$ \textit{Member, IEEE}\\ 
$^{1,2,3}$M S Ramaiah Institute of Technology, Bangalore, India; $^4$Infineon Technologies}
E-mail: chalukya.bhat@gmail.com$^1$, aniruddh.herle@gmail.com$^2$, bcjanmay.edu@gmail.com$^3$,\\ swaraj.kali@gmail.com $^4$}
\maketitle
\begin{abstract} 
Internet of Things (IoT) applications have opened up numerous possibilities to improve our lives. Most of the remote devices, part of the IoT network, such as smartphones, data loggers and wireless sensors are battery powered. It is vital to collect battery measurement data (Voltage or State-of-Charge (SOC)) from these remote devices. Presence of noise in these measurements restricts effective utilization of this dataset. This paper presents the application of Unscented Kalman Filter (UKF) to mitigate measurement noise in smartphone dataset. The simplicity of this technique makes it a constructive approach for noise removal. The datasets obtained after noise removal could be used to improve data-driven time series forecasting models which aid to accurately estimate critical battery parameters such as SOC. UKF was tested on noisy charge and discharge dataset of a smartphone. An overall Mean Squared Error (MSE) of 0.0017 and 0.0010 was obtained for Voltage charge and discharge measurements. MSE for SOC charge and discharge data measurements were 0.0018 and 0.0010 respectively.
\end{abstract}
 
\IEEEpeerreviewmaketitle

\vspace*{0.25cm}


 \section{Introduction} 

\begin{figure}[htbp!]
\centering
\includegraphics[scale=0.46]{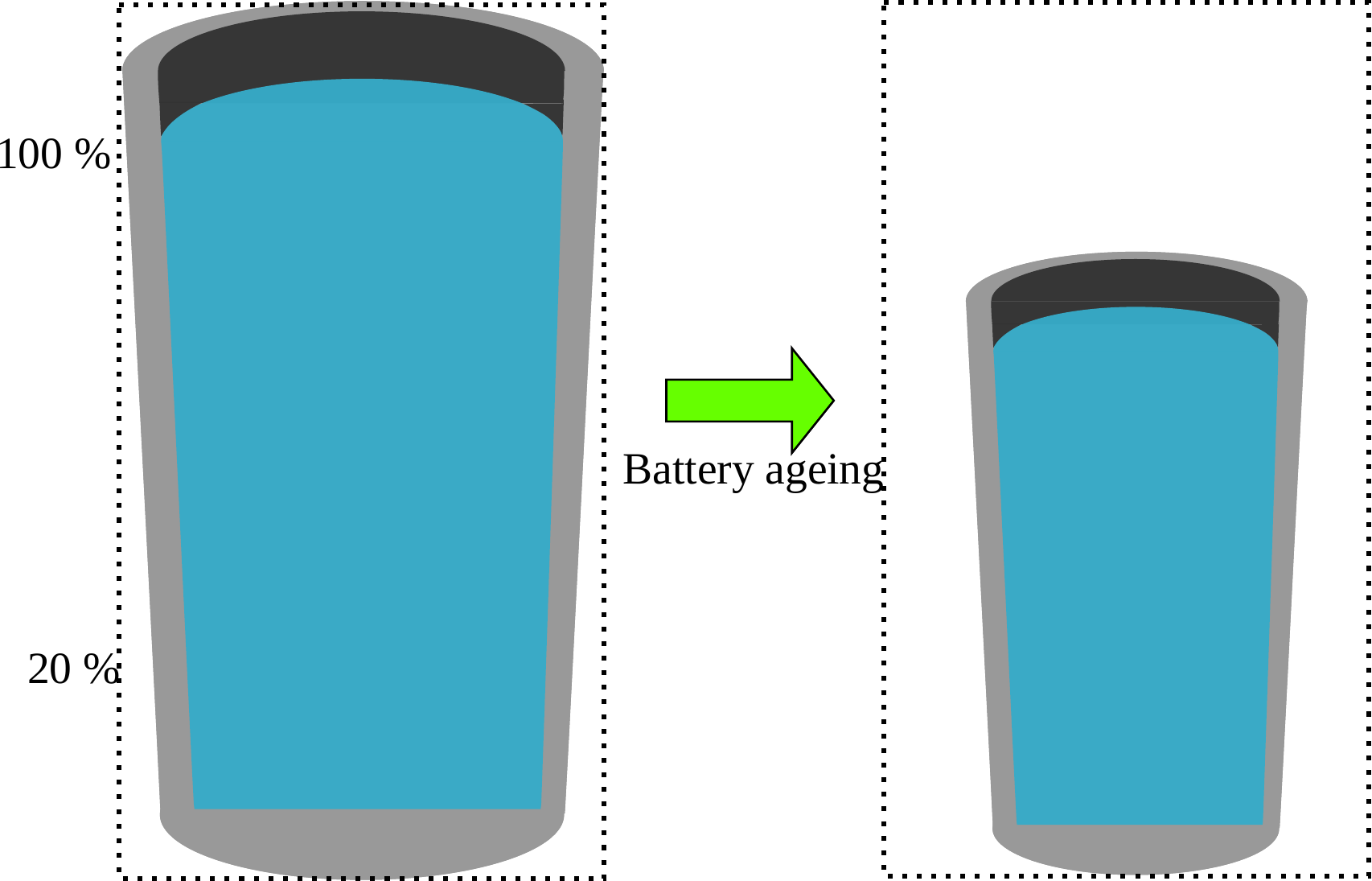}
\caption{Battery capacity fade overtime}
\label{fig: capacity fade}
\end{figure}

\begin{figure}[htbp!]
\centering
\includegraphics[scale=0.8]{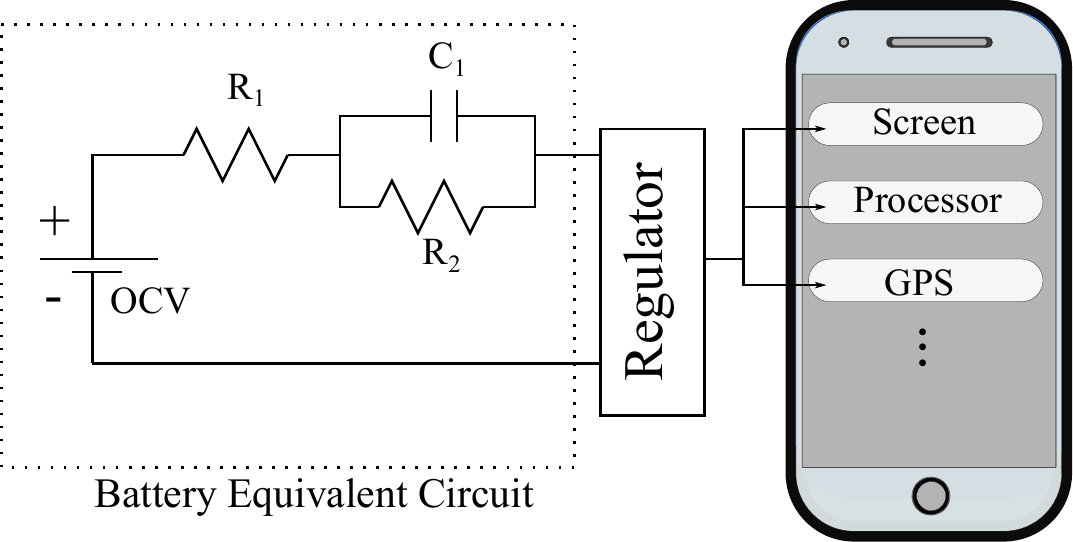}
\caption{Smartphone Battery Equivalent Circuit}
\label{fig:Ph Batt}
\end{figure}

Increase in number of IoT enabled devices have led to significant improvements in home automation \cite{1}. Remote devices are primarily powered by Lithium-ion based batteries. Advancements in energy storage systems have not kept pace with growth of IoT devices \cite{2}. Smartphones, one of  the most commonly available IoT device, over the past decade have become a ubiquitous part of our daily lives. Unanticipated smartphone shutdown have hampered the device usage. Accurate State-of-Charge (SOC) estimation aids to predict the available usage time of a smartphone \cite{3}.\\
\indent SOC estimation of batteries is challenging due to the inherent non-linear battery characteristics. Added to this non-linearity, batteries suffer from capacity fade, the collective capacity of the battery reduces overtime, this reduction is illustrated in Fig \ref{fig: capacity fade}. Public battery datasets are available \cite{nasa}, \cite{pana}, but are collected in highly controlled environments. Collection of smartphone battery data is needed to evaluate and anticipate the approximate time-limit for device shutdown. \\
\indent Presence of noise in such datasets is a cause for concern as they negatively impact the inference obtained from data-driven models \cite{noise1} - \cite{noise3}. Data cleaning and curation is a vital pre-processing task in all data-driven techniques \cite{dataclean}. \\
\indent There have been multiple research endeavors to use UKF for SOC estimation \cite{CDKF} - \cite{kf2}. This paper is the first to address the issue of smartphone battery noise removal using UKF. For the sake of completeness the equivalent circuit model used is illustrated in Fig \ref{fig:Ph Batt} \cite{lu}. Section II of this paper provides details about UKF followed by discussion of results in Section III. Conclusion of the present work is presented in Section IV.

\section{Unscented kalman filter}

Unscented kalman filter employs a nonlinear unscented transformation (UT), UKF performs estimation by iterative update in the mean and error covariance. UKF is advantageous for estimating the properties of a random variable subjected to  nonlinear transformation. UKF is helpful for noise reduction in the state variables of non-linear functions by adopting initial values signal under consideration and evaluating future values iteratively. The State and Measurement Equations are provided below:

$$
\begin{array}{l}
\mathbf{x}_{k}=\mathbf{f}\left(\mathbf{x}_{k-1}\right)+\mathbf{w}_{k-1} \\
\mathbf{z}_{k}=\mathbf{h}\left(\mathbf{x}_{k}\right)+\mathbf{v}_{k}
\end{array}
$$
In the equation $\mathbf{x}_{(k-1)}$ is the state vector at the moment $k-1$ and $\mathbf{x}_{k}$ is the state vector at the instance $k$.   $\mathbf{w}_{k-1}$ is the process noise at $k-1$ and $\mathbf{v}_{k}$ is the measurement noise at $k$. $\mathbf{f}(\mathbf{x}_{k-1})$ and $\mathbf{h}\left(\mathbf{x}_{k}\right)$ are non-linear functions \cite{kt1} - \cite{kt3}. 
The procedure followed for noise removal is given below:

\textbf{Step 1}:The \textbf{State equations} of the terminal Voltage $E_T$ and state-of-charge $SOC$ are given as here:
\begin{equation}
\begin{split}
E_{T(t+1)} = E_{T(t)} + \frac{V_1}{R_1(SOC,T_b)*C_1(SOC,T_b)} \hspace*{0.1cm} - \\
\frac{I}{C_1(SOC,T_b)} + V_t
\end{split}
\end{equation}
\begin{equation}
\begin{split}
S O C_{T+1} = \mathrm{SOC}_{T} + -\frac{I}{3600 \cdot C_{q}} T_s + V_{t}
\end{split}
\end{equation}

Where $V_t$ is the process noise, ${T_s}$ is time step which is set to 1 second and $C_q$ is the capacity of the smartphone battery.
The \textbf{Measurement equations} used to determine clean terminal voltage and state of charge are:

\begin{equation}
\widehat{E_{T(t)}} = E_{T(t)} + W_t
\end{equation}
\begin{equation}
\widehat{SOC_{(t)}} = SOC_{(t)} + W_t
\end{equation}
The symbols used have been defined below:
\begin{itemize}
\item $W_t$ = Measurement Noise 
\item $R_1$ = Equivalent battery resistance 
\item $V_1$ = Voltage across $R$ and $C$ as seen in Fig \ref{fig:ECM}
\item $V_1$ and $SOC$ are preliminary values
\item Current $I$ and battery temperature $T_b$ are obtained from logged data
\end{itemize}

\textbf{Step 2}: White noise of value $10^{-4}$ $\frac{W}{Hz}$ is added to the measured value of terminal voltage ($E_T$).

\textbf{Step 3}: UKF parameters including the initial covariences are defined as provided in Table \ref{table UKF}.

\textbf{Step 4}: Compare UKF output and with the dataset and calculate Mean Square Error (MSE). The block diagram of the approach is highlighted in Fig \ref{fig:UKF Block}.

\subsection{Battery Equivalent Circuit}

\begin{figure}[htbp!]
\centering
\includegraphics[scale=0.9]{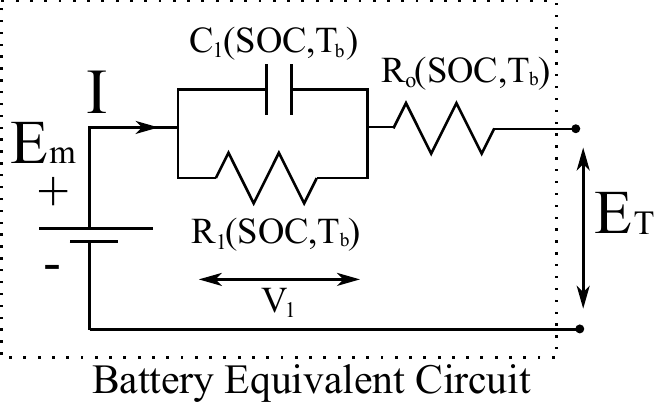}
\caption{Battery Equivalent Circuit illustrating passive elements}
\label{fig:ECM}
\end{figure}

Battery Equivalent Circuit model is provided in Fig \ref{fig:ECM}. Passive Battery parameters $R_1$ and $C_1$ are a function of State of Charge and battery and temperature. To improve predictions of UKF thermal effects of the battery are also considered. The resistance and capacitance values at temperatures 278K, 293K and 313K are provided in table \ref{table R} and table \ref{table C} \cite{lu}. 

\begin{table}[t!]
\begin{center}
\caption{Table Providing resistance $R_1$ values at temperatures 278 K, 293 K and 313 K for variations in SOC} 
\label{table R}
\begin{tabular}{ l c c c} 
 \hline
 \textbf{SOC} & \textbf{$R_{278 K}$ $\Omega$} & \textbf{$R_{293 K}$ $\Omega$} & \textbf{$R_{313 K}$ $\Omega$} \\
\hline\
0 &  0.0109 &  0.0029 &  0.0013 \\
0.1 &  0.0069 &  0.0024 &  0.0012 \\
0.25 &  0.0047 &  0.0026 &  0.0013 \\
0.5 & 0.0034 &  0.0016 &  0.001 \\
0.75 &  0.0033 &  0.0023 &  0.0014 \\
0.9 &  0.0033 &  0.0018 &  0.0011 \\
1 &  0.0028 &  0.0017 & 0.0011\\
 \hline
\end{tabular}
\end{center}
\end{table}
\begin{table}[t!]
\begin{center}
\caption{Table Providing Capacitance $C_1$ values at temperatures 278 K, 293 K and 313 K for variations in SOC} 
\label{table C}
\begin{tabular}{ l c c c} 
 \hline
 \textbf{SOC} & \textbf{$C_{278 K}$ $\mu F$} & \textbf{$C_{293 K}$ $\mu F$} & \textbf{$C_{313 K}$ $\mu F$} \\
\hline\
0 &  1913.6 &  12447 &  30609 \\
0.1 &  4625.7 &  18872 &  32995 \\
0.25 & 23306 &  40764 &  47535 \\
0.5 & 10736 &  18721 &  26325 \\
0.75 &  18036 & 33630 & 48274 \\
0.9 &  12251 &  18360 & 26839 \\
1 &  9022.9 & 23394 &  30606 \\ 
 \hline
\end{tabular}
\end{center}
\end{table}

\section{Results and Discussion}

\begin{figure}[htbp!]
\centering
\includegraphics[scale=0.6]{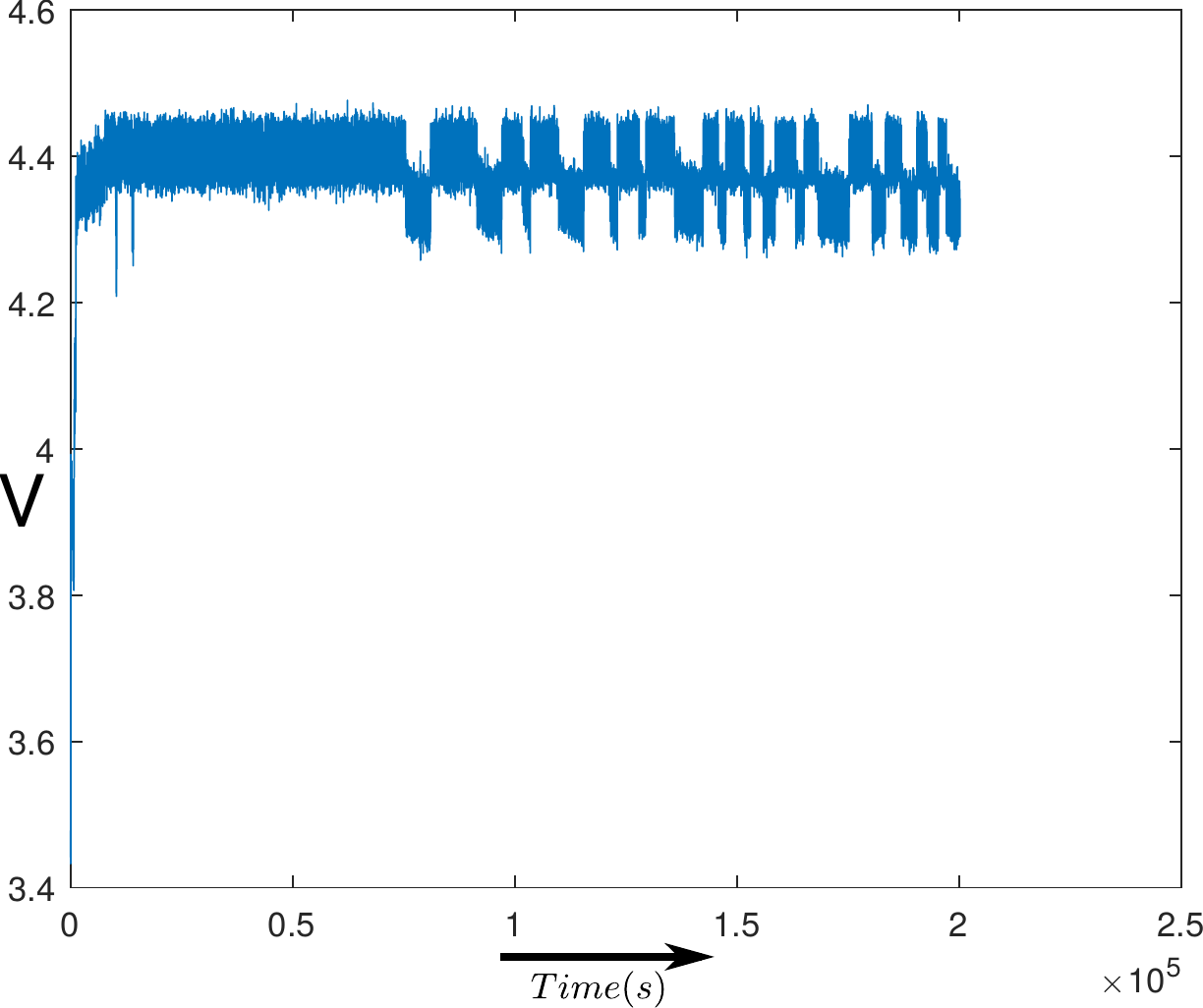}
\caption{Presence of noise in battery voltage during charging}
\label{fig: V charge}
\end{figure}

\begin{figure}[htbp!]
\centering
\includegraphics[scale=0.6]{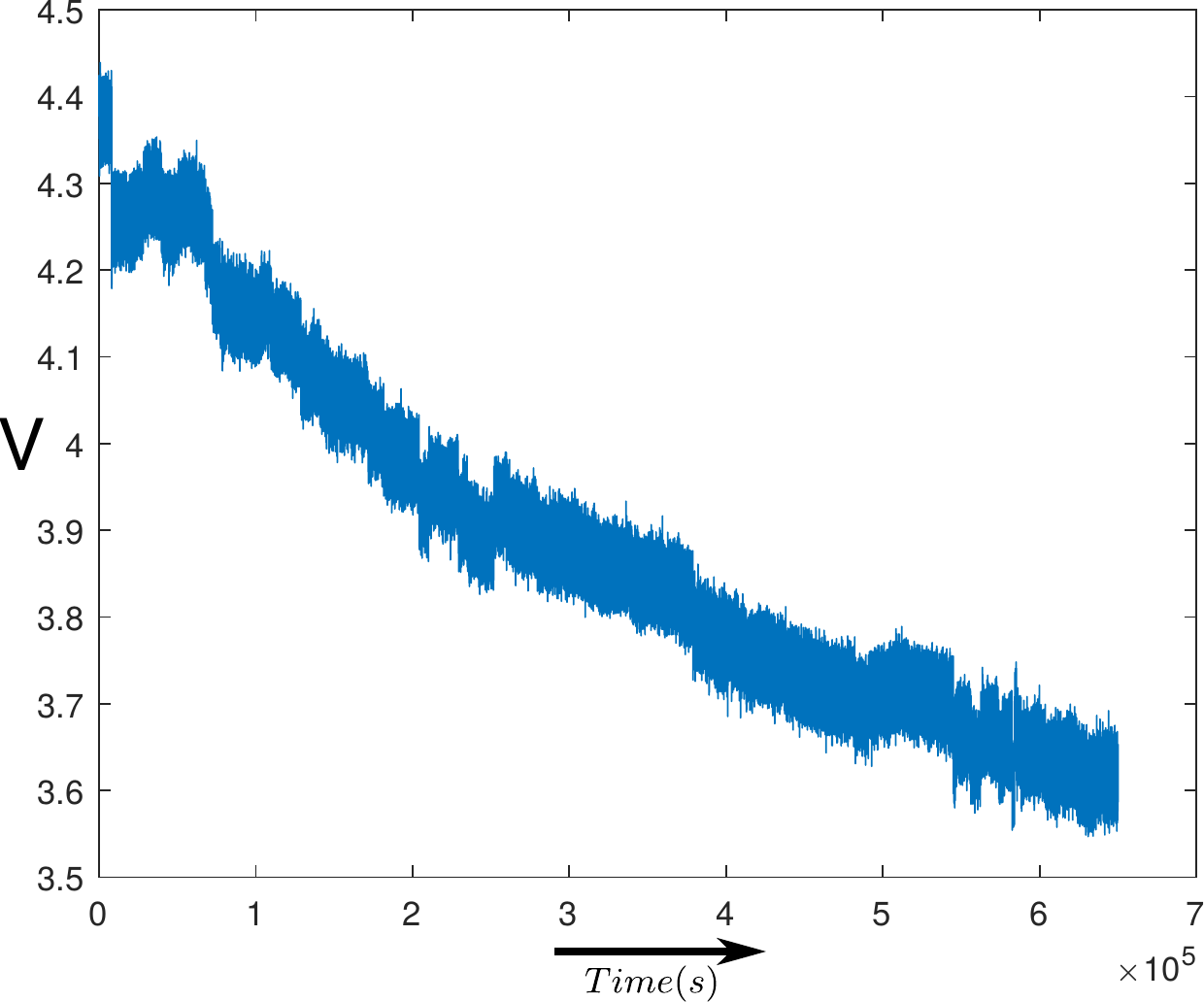}
\caption{Presence of noise in battery voltage during discharge}
\label{fig: V dis}
\end{figure}

\begin{figure}[htbp!]
\centering
\includegraphics[scale=0.6]{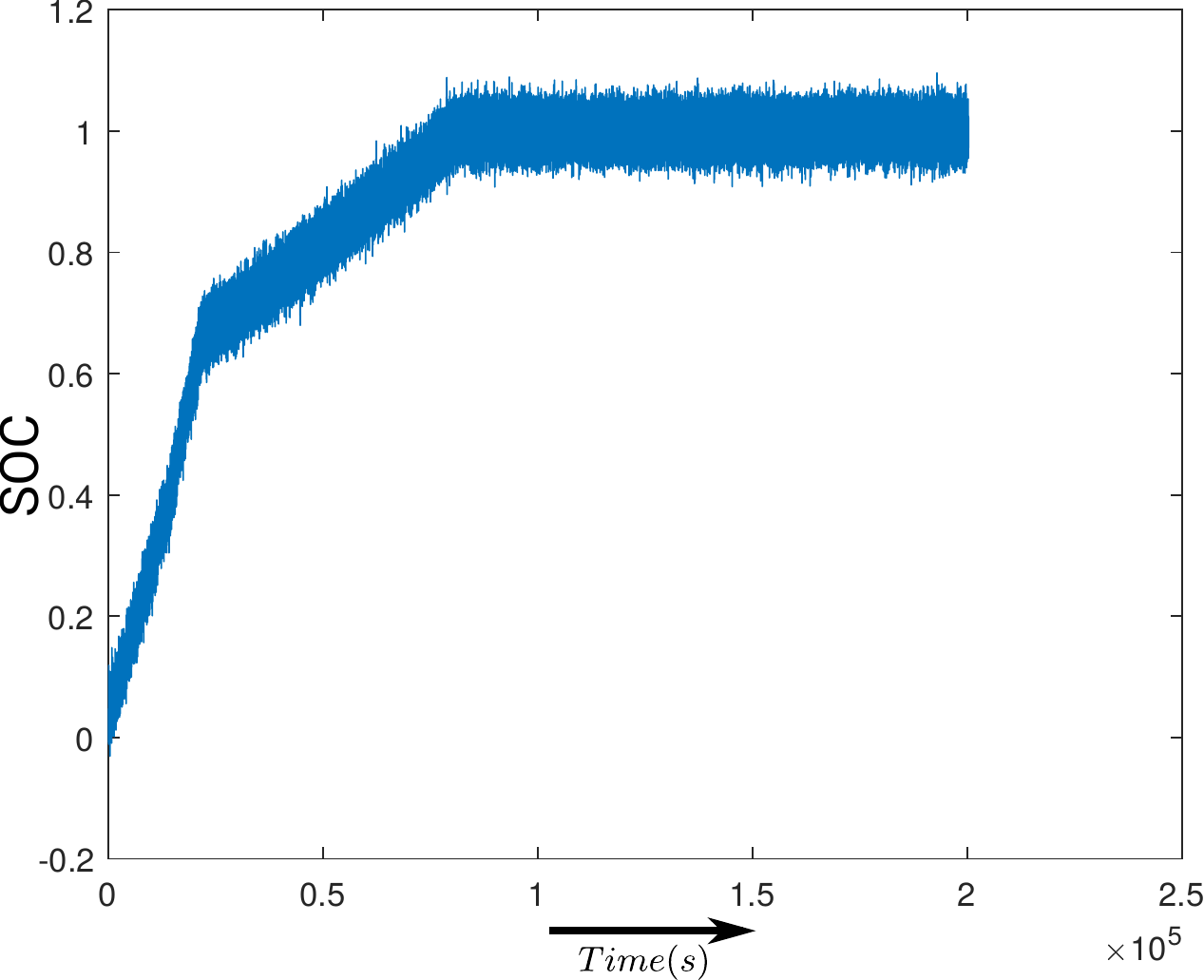}
\caption{Presence of noise in battery SOC during charging}
\label{fig: SOC charge}
\end{figure}

\begin{figure}[htbp!]
\centering
\includegraphics[scale=0.6]{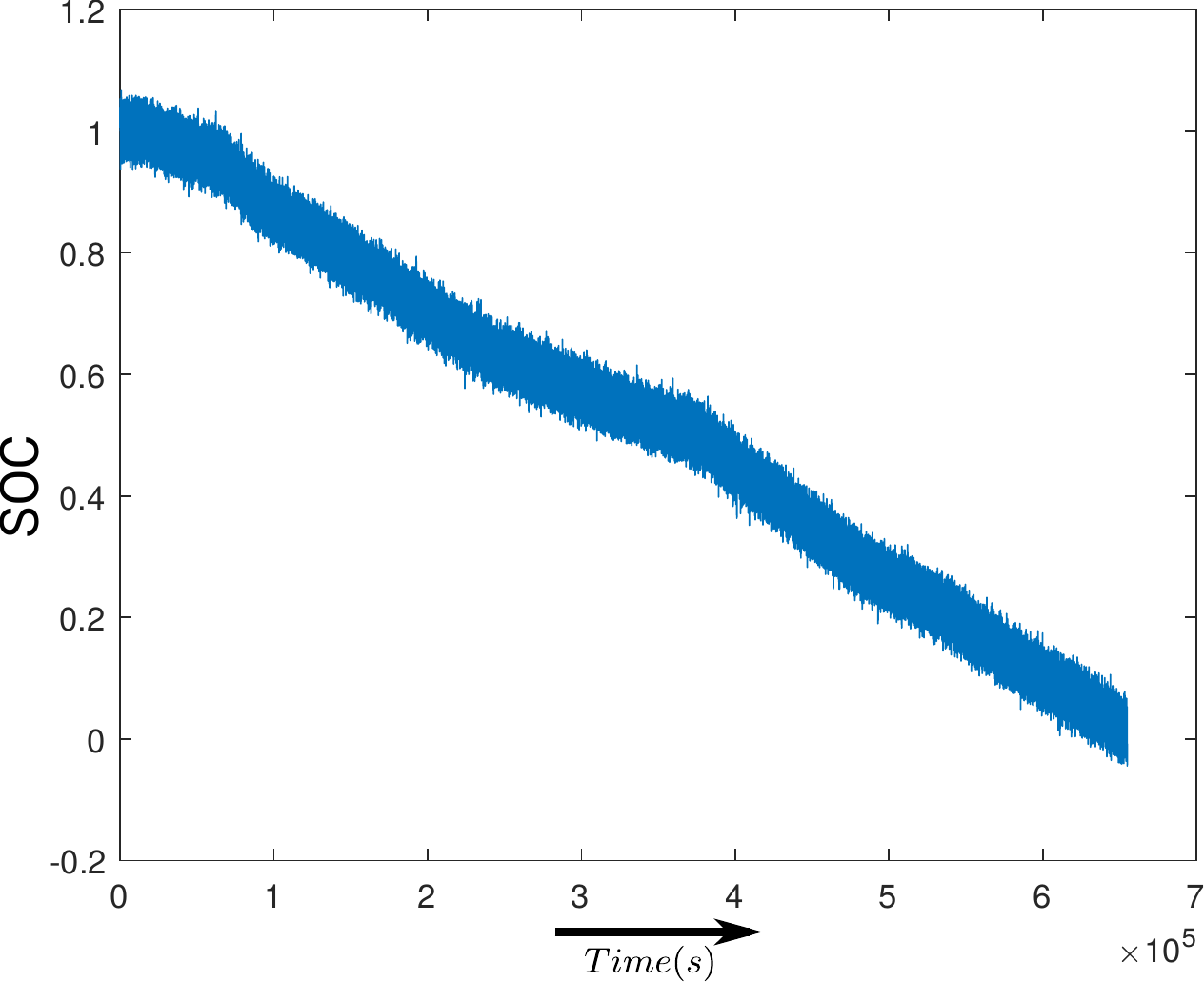}
\caption{Presence of noise in battery SOC during discharge}
\label{fig: SOC dis}
\end{figure}

\begin{figure}[htbp!]
\centering
\includegraphics[scale=0.18]{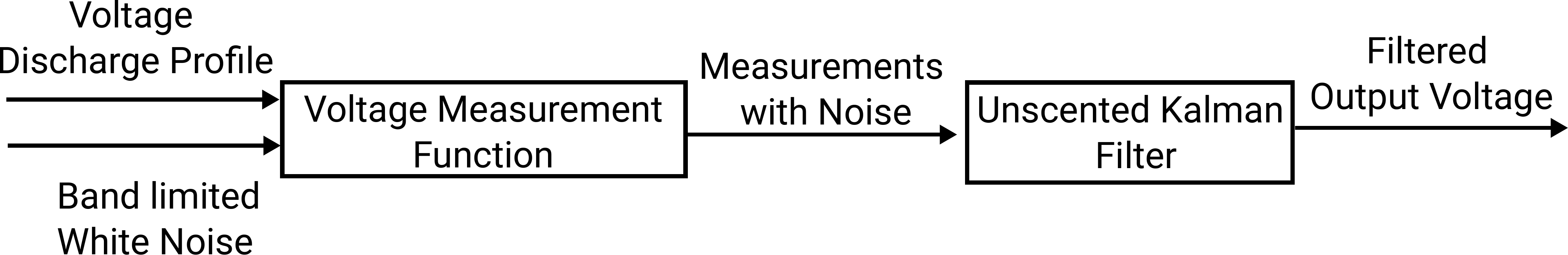}
\caption{Block Diagram of UKF}
\label{fig:UKF Block}
\end{figure}

\begin{figure*}[t]
\centering
\includegraphics[scale=0.48]{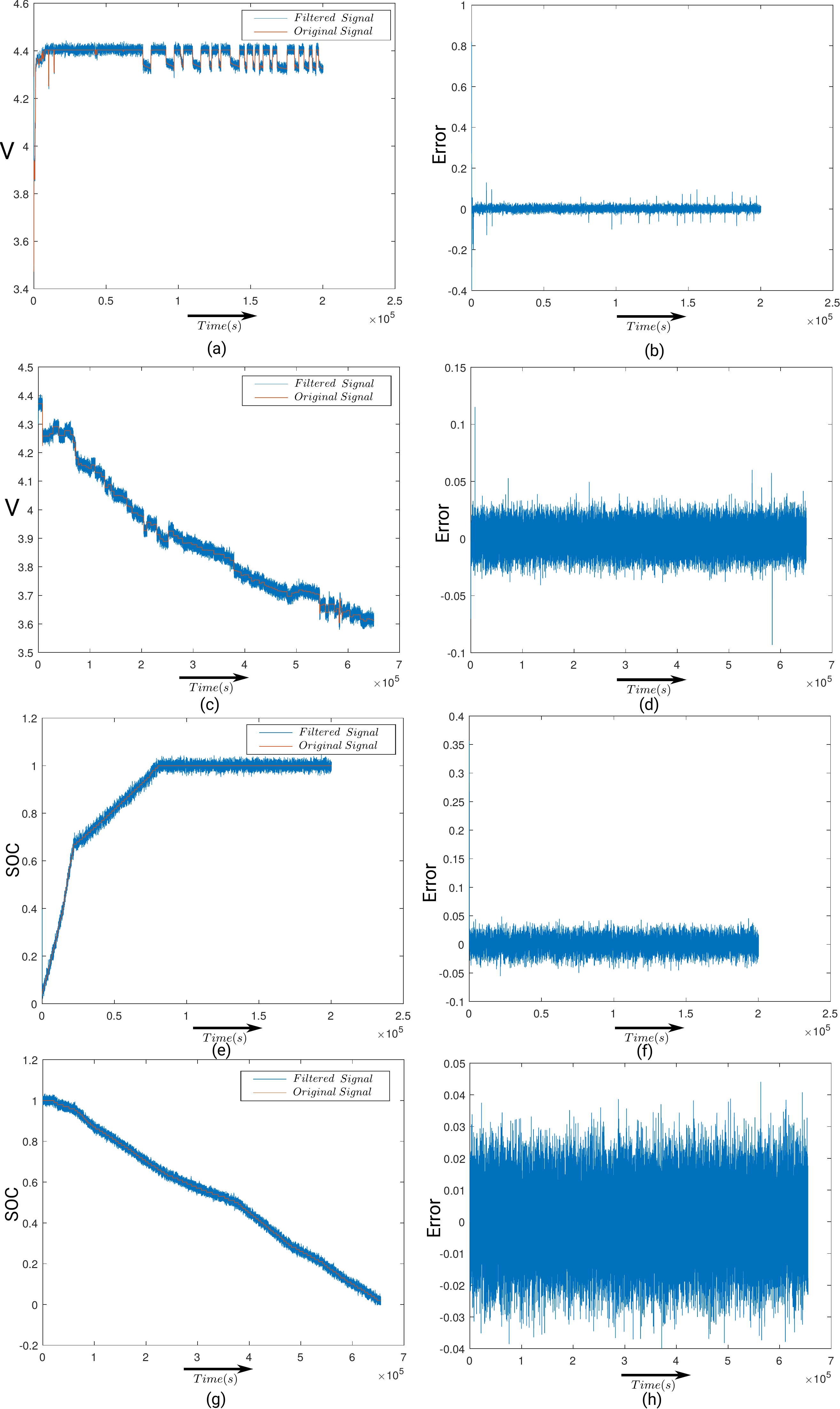}
\caption{Output of UKF for charging voltage characteristics is shown in (a) the error is shown in (b). The discharging characteristics of the battery are compared with ground truth values in (c) and the error is illustrated in (d). Ground truth values for State-of-charge (SOC) for charging and discharging is shown in (e) and (g). Error plots for SOC are depicted in (f) and (h) respectively. Total simulation time was 2500 seconds with a simulation time step of 0.1 second for battery charging dataset \& 7000 seconds with a 0.1 second during discharging}
\label{fig:all}
\end{figure*}

All measurements with an added White Noise of value $10^{-4}$ $\frac{W}{Hz}$ are shown in Fig \ref{fig: V charge}, Fig \ref{fig: V dis}, Fig \ref{fig: SOC charge} and Fig \ref{fig: SOC dis} respectively. The block diagram displaying noise addition is in Fig \ref{fig:UKF Block}. Preliminary error rate was assumed to be 10 \% of terminal voltage. All the other smartphone battery parameters are taken from the datasheet \cite{batt}. The filtered output voltage has been compared with ground truth values as shown in Fig \ref{fig:all}.
 
\begin{table}[hbpt!]
\begin{center}
\caption{UKF parameters} 
\label{table UKF}
\begin{tabular}{ l c} 
 \hline
 \textbf{Parameter} & \textbf{Value} \\
\hline
 Process Noise Covariance & 0.5\\
 Initial State Voltage (charge) & 2.5 V \\
 Initial State Voltage (discharge) & 4.2 V \\
 Initial state of charge (charge) & 0.01 \\
 Initial state of charge (discharge) & 1 \\
 Initial state Covariance (voltage) & 0.01 \\
 Initial state Covariance (SOC) & 0.01 \\
 Measurement Noise Covariance & 1 \\
Battery Capacity ($C_q$) & 4000mAh \\ 
 
  \hline
\end{tabular}
\end{center}
\end{table}

\subsection{Mean Squared Error}

To assess the standard of UKF output, Mean Squared Error (MSE) was used. MSE provides the average squared difference between the estimated and the ground truth voltage value. The number of data points selected for this study was 25000 and 70000 for charging and discharging respectively.
\begin{equation}
MSE = \frac{1}{n} \sum_{i=1}^{n}\left(Y_{i}-\hat{Y}_{i}\right)^{2}
\end{equation}

\begin{itemize}
\item MSE = Mean Squared Error
\item n = number of data points
\item $Y_i$ = Observed Values
\item $\hat{Y}_{i}$ = Predicted Values
\end{itemize}


\section{Conclusion}
Removal of noise in measured data is important to extract maximum value from the dataset. This paper explored the use of Unscented Kalman Filter (UKF) to mitigate noise in measured data. The voltage \& State-of-Charge measurements were selected to test the proposed approach. This technique does not need extensive computational capabilities and can be deployed in most smartphones with ease. UKF resulted in a Mean Squared Error (MSE) of 0.0017 and 0.0010 for charge and discharge measurements of smartphone battery voltage. A MSE of 0.0018 and 0.0010 was obtained for SOC measurements for charge \& discharge conditions.

\clearpage


\clearpage



\end{document}